\documentstyle{article}
\newcommand{\citetie}{$^-$}

\begin{document}

\ \

\vspace{1.3cm}

\begin{center}

{\large\bf RECYCLE OF RANDOM SEQUENCES}

\bigskip
\bigskip

{NOBUYASU ITO}

\medskip

{\it
Institut f\"ur Theoretische Physik, Universit\"at zu K\"oln \\
       D-W-5000 K\"oln 41, Germany \\
and \\
Computing and Information Systems Center\\
Japan Atomic Energy Research Institute \\
Tokai, Ibaraki 319-11, Japan \\
}

\medskip

\medskip

{MACOTO KIKUCHI}

\medskip

{\it
Institut f\"ur Physik, Johanness-Gutenberg Universit\"at \\
and Max-Planck Institut f\"ur Polymerforschung \\
D-W-6500 Mainz, Germany\\
and\\
Department of Physics, Osaka University\\
Toyonaka, Osaka 560, Japan\\
}

\medskip

and

\medskip

{YUTAKA OKABE}

\medskip

{\it
Department of Physics, Tokyo Metropolitan University\\
Tokyo 192-03, Japan\\
}

\bigskip
\bigskip

Received 6 November 1992

\bigskip
\bigskip

\parbox{11.5cm}{
\noindent
The correlation between a random sequence and its transformed
sequences is studied.
In the case of a permutation operation or, in other word, the
shuffling operation, it is shown
that the correlation can be so small that
the sequences can be regarded as independent random sequences.
The applications to the Monte Carlo simulations are also given.
This method is especially useful in the Ising
Monte Carlo simulation.

\bigskip

{\it Keywords}: random number; Monte Carlo simulation; Ising model.

}

\end{center}

\medskip

\pagebreak

\section{Introduction}

Random numbers are indispensable to the Monte Carlo simulation, which
is one of the most important applications of supercomputers.
The recent development of vector computers makes more accurate
Monte Carlo simulation possible.
The Ising Monte Carlo simulation is the most highly developed application
and the precision\cite{NISMA}\citetie\cite{BGHP92} now reaches to the order of
$10^{-6}$ and sometimes
such simulations consume of the order of $10^{14}$ random numbers.

When a high-precision Monte Carlo simulation is made, two points are
important.
One is the use of correct pseudo-random number.
It is a quite reasonable requirement.
And the other is the efficient use of random numbers.
For example, the fastest Monte Carlo algorithm using the single-spin
updating dynamics of the Ising model\cite{BDS86}\citetie\cite{KIK92} uses
one random number for updating several spins which are
coded in a computer
word({\it independent-system coding
technique})\cite{BDS86}\citetie\cite{NIYK90}.
The spins in a word belong to physically independent systems and therefore
it is expected that such a technique does not make a problem.
The validity of this algorithm relies on the physical independence of
the simulated systems.
When we want to know the expectation values at many different temperatures or
fields, this kind of efficient use of random number is very helpful
and it has been used in many studies\cite{NISMA,NISMB,KO86}\citetie\cite{KO92}.
But if we want to know the expectation value at one temperature,
this algorithm was not useful.
To overcome this difficulty, the random shuffling of the
Boltzmann factor tables(BFT, see Section 6) has been proposed\cite{CM86}.
Although it seemed to work,
the validity and/or the limitation of such an operation has not been
made clear.

If we transform a given sequence of random numbers,
$\{ r_n $; $n=0$, $1$, $2$, $3$, $\cdots \}$, using some map $f$ from
interval $I=[0$, $1]$ to $I$, how does the new sequence
$\{ f(r_n) $; $n=0$, $1$, $2$, $3$, $\cdots \}$ behave?
If $f(x)=x^2$ or some other smooth analytic function is used,
the sequences $\{ r_n\}$ and $\{ f(r_n)\}$ will be strongly correlated.
Intuitively if we use $f(x)$ which maps any real number in $I$ into $I$
randomly, the sequence $\{ r_n \}$ and $\{ f(r_n)\}$ may be
independent random sequences.
On the digital computer, the reliable random real numbers
distributed between $0$ and $1$ are usually generated by
normalizing random integers which are generated by a recursion
relation\cite{TAUS65}\citetie\cite{IK90}.
When the random integer is distributed uniformly between $0$ and $N-1$,
the above mentioned map $f$ can be regarded as a map from
$I_N=\{ 0$, $1$, $2$, $\cdots$, $N-1\}$ to $I_N$.
This kind of discrete random number is treated in this paper.

The purpose of the present paper is to study the behavior of
random sequences under transformations and to
clarify whether there is room for improvement
in Monte Carlo simulation in terms of this transformation strategy.
And a new algorithm named {\it recycle algorithm} is proposed.
This algorithm makes the meaning, validity and limitation
of the BFT shuffling operation clear.

Some relevant lemmas about the random sequence transformation are given
in the next section.
The algorithm of random permutation generation which is important in the
application of our algorithm is studied in the third section.
The general study of the efficiency of our algorithm is
made in the fourth section.
In the fifth section, a simple application to the static
Monte Carlo integration is shown.
The application to the dynamic Monte Carlo simulation of the Ising
model is given in the sixth section.
The final section is for the conclusion and remarks.

\section{Some Lemmas}

In this section, some useful lemmas are given. Their proofs are not
given but they are easily proved.

\bigskip

\noindent {\bf Notation }

$K$ is a field( for example, real field or complex field)
and $\Omega$ is any set with $N$ elements, where $N$ is a positive integer.
$T_N$ denotes the set of all the maps from $\Omega$ to $\Omega$.
There are $N^N$ maps in $T_N$.
The subset of $T_N$ made of injections are denoted by $S_N$,
which is naturally regarded as the symmetric group of order $N$.
$F^N_M$ denotes the set of all the maps from $\Omega^M$ to $K$.
$f\circ t$ denotes the composition of $t\in T_N$ and $f\in F^N_M$.
In the following lemmas, $N$ is assumed to be larger than $2M$.

\bigskip

Although the lemmas below are proved under above-mentioned general
situations, we are interested in the application to the random sequence
and Monte Carlo simulations.
In these cases, the real field is usually used
as $K$. And $\Omega$ corresponds to the possible values of
random number, that is,
\begin{equation}
\Omega =\{ {i\over N}; i=0, 1, 2, \cdots, N-1 \} .
\end{equation}
Instead of continuous random numbers, discrete random numbers are
considered and it is appropriate because the pseudo-random number
generated by a digital computer usually takes such discrete values.

\bigskip

\noindent {\bf Definition }

For any $f\in F^N_M$, the average of $f$, $I^N_M[f]$, is defined by
\begin{equation}
I^N_M [f] =
  {1\over N^M}
    \sum_{\omega_1, \omega_2, \omega_3, \cdots, \omega_M\in\Omega}
       f(\omega_1, \omega_2, \omega_3, \cdots, \omega_M).
\end{equation}

\bigskip

\noindent {\bf Lemma~1. }

If $I^N_1[f\circ t] = I^N_1[f]$ for a given $t\in T_N$ and any $f\in F^N_1$,
$t$ is an element of $S_N$. Inversely, if $t$ is an element of $S_N$,
$I^N_1[f\circ t] = I^N_1[f]$ for any $f$ in $F^N_1$.

\bigskip

This Lemma~1 suggests that the permutation operation is better than
other operations in $T_N$ to transform the random sequences into
other random sequences.
If a permutation is used, the averaged value of $f$ is not biased
at all.
If we use the permutation, there is no deviation of this kind
even for $M$ variable function:

\bigskip

\noindent {\bf Lemma~2. }

For any $s\in S_N$ and any $f\in F^N_M$,
\begin{equation}
I^N_M [f\circ s] = I^N_M[f].
\end{equation}
Therefore
\begin{equation}
<\Delta I^N_M[f\circ s]^2>_{s\in S_N} =0,
\end{equation}
where $<\cdot >_{a\in A}$ denotes the average over a set $A$
and
\begin{equation}
<\Delta I^N_M[f\circ s]^2 >_{s\in S_N}
= <I^N_M[f\circ s]^2>_{s\in S_N} - <I^N_M [f\circ s]>_{s\in S_N}^2.
\end{equation}

\bigskip

But if an other transformation is used, a deviation of
the order of $1/\sqrt{N}$ appears as shown in the following Lemma~3:

\bigskip

\noindent {\bf Lemma 3. }

For any $f\in F^N_M$,
\begin{equation}
<I^N_M [f\circ t]>_{t\in T_N} = I^N_M[f]
\end{equation}
and
\begin{equation}
<\Delta I^N_M[f\circ t]^2 >_{t\in T_N} = {1\over N} I^N_M[\Delta f^2],
\end{equation}
where
\begin{equation}
I^N_M[\Delta f^2] = I^N_M[f^2] - I^N_M[f]^2.
\end{equation}

\bigskip

The following Lemma~4 shows that the correlation between a random sequence and
the sequences obtained by its permutation becomes small when the
resolution of the random number $1/N$ becomes high.

\bigskip

\noindent {\bf Lemma~4. }

For any $s\in S_N$ and any $f$, $g\in F^N_M$,
\begin{equation}
<I^N_M [f\cdot g\circ s]>_{s\in S_N} -I^N_M[f] I^N_M[g]
= {C_{f,g}\over N}+O({1\over N^2})
\end{equation}
and
\begin{equation}
<\Delta I^N_M [f\cdot g\circ s]^2>_{s\in S_N}
={E_{f,g}\over N} +O({1\over N}),
\end{equation}
where
\begin{equation}
C_{f,g} =
\sum_{i > j}
(I^N_M[f]-D^N_{M,i,j}[f])
(I^N_M[g]-D^N_{M,i,j}[g]),
\end{equation}
\begin{equation}
E_{f,g} =
\sum_{i=1}^M \sum_{j=1}^M
(I^N_M[f]^2-W^N_{M,i,j}[f])
(I^N_M[g]^2-W^N_{M,i,j}[g]),
\end{equation}
\begin{equation}
  D^N_{M,i,j}[f]
= {1\over N^{M-1}} \sum_{\omega_i=\omega_j}
    f(\omega_1, \omega_2, \omega_3, \cdots, \omega_M)
\label{DEFD}
\end{equation}
and
\begin{equation}
  W^N_{M,i,j}[f]
= {1\over N^{2M-1}} \sum_{\omega_i=\omega_j'}
    f(\omega_1, \omega_2, \omega_3, \cdots, \omega_M)
    f(\omega_1', \omega_2', \omega_3', \cdots, \omega_M').
\label{DEFW}
\end{equation}
The summation in eq.~(\ref{DEFD}) runs over all the
elements $(\omega_1$, $\omega_2$, $\omega_3$, $\cdots$, $\omega_M)\in \Omega^M$
which satisfy the conditions that $\omega_i=\omega_j$.
The summation in eq.~(\ref{DEFW}) runs over all the
elements $(\omega_1$, $\omega_2$, $\omega_3$, $\cdots$, $\omega_M)\in \Omega^M$
and $(\omega_1'$, $\omega_2'$, $\omega_3'$, $\cdots$, $\omega_M')\in \Omega^M$
which satisfy the conditions that $\omega_i=\omega_j'$.
\bigskip

The lemma~1, 2 and 4 suggests a new algorithm of random number generation
and usage:

\bigskip

\noindent {\bf Algorithm(Recycle Use of Random Number)}

{}From a sequence of random integers, a new random sequence is generated
by applying permutation operations which are selected randomly.

\bigskip

The efficiency analysis and real implementations of this algorithm
are given in the rest sections.

Here one remark should be made. When a fixed permutation $s\in S_N$ is
applied to a random number repetitively, the randomness of the
obtained sequence is worse than the sequence obtained by the above
algorithm as is shown in the Lemma~5:

\smallskip

\noindent {\bf Definition }

For any $f\in F^N_1$, $s\in S_N$ and $\omega\in\Omega$,
$J^N$ is defined by
\begin{equation}
J^N[f; s, \omega ]
=\lim_{L\rightarrow\infty}{1\over L}\sum_{k=0}^{L-1} (f\circ s^k)(\omega ).
\end{equation}

\bigskip

\noindent {\bf Lemma~5. }

For any $f\in F^N_M$ and $\omega\in\Omega$,
\begin{equation}
<J^N [f;s,\omega ]>_{s\in S_N}
= I^N_1[f] + {\rho_N \over N} (f(\omega ) -I_N[f]) +O({1\over N}),
\end{equation}
\begin{equation}
<J^N [f;s,\omega ]>_{s\in S_N, \omega\in\Omega} = I^N_1[f]
\end{equation}
and
\begin{equation}
<\Delta J^N[f;s,\omega ]^2>_{s\in S_N, \omega\in\Omega}
= {\rho_N\over N} I^N_1 [\Delta f^2] +O({1\over N}),
\end{equation}
where
\begin{equation}
\rho_N = \sum_{k=1}^{N} {1\over k}.
\end{equation}

\smallskip

Note that the $\rho_N$ behaves as $C_{\rm E}+\log N$ when $N$ is large,
where $C_{\rm E}$ is Euler's constant. Therefore the deviation is
the order of $\sqrt{\log N/N}$.

\section{Permutation Generation}

In the previous section, it is shown that a random sequence and its
transformed sequence by permutation are {\it almost} independent with
each other.
Now the question is whether this recycle algorithm of a random sequence
improves the efficiency of the real simulation or not.

In the real application, the random integers uniformly distributed
between $0$
to $2^n-1$ are easily generated and such a random integer is
to be transformed by one table-lookup.
For example, such random integers are generated by Tausworthe's
method\cite{TAUS65}\citetie\cite{IK90}.
The table holds a permutation of integers between $0$ to $2^n-1$.
When $20$ bit random integer is used, that is, $n=20$, the necessary
memory storage for one such table is $4$ byte $\times 2^{20} = 4$ Mbyte.
This is not a large table on the modern computers.

The use of discrete random integers instead of real random numbers
causes a correlation of the order of $2^{-n/2}$ as it is shown
in Lemma 4.
Therefore we have to analyze the efficiency carefully.
For that purpose, it is necessary to know how to generate a
permutation randomly.
Random numbers are necessary to prepare a permutation randomly and
it might be possible that the total amount of necessary random numbers
was larger.
Therefore the permutation generation methods are studied in this section
to clarify the efficiency of the recycle algorithm.

Consider $N$ objects which are
in the boxes, $A_0$, $A_1$, $A_2$, $\cdots$, $A_{N-2}$ and $A_{N-1}$
at the beginning.
Now we want to shuffle the objects and generate a new order which is
independent of the original order. This is called the {\it shuffling
problem}.

This problem is easily solved by the following algorithm:
Firstly, all the objects are put out of the boxes.  Then select one object
randomly
and put it in $A_0$. And then select one object from the remains randomly and
put it in $A_1$ and so on.
An naive implementation is the following FORTRAN subroutine:
\begin{verbatim}
      SUBROUTINE PERM1(N,ILIST)
      IMPLICIT REAL*8 (A-H,O-Z)
      DIMENSION ILIST(0:N-1)
      DO 10 I=N-1,0,-1
        J=INT(RANDOM()*DFLOAT(I+1))
        ITMP=ILIST(I)
        ILIST(I)=ILIST(J)
        ILIST(J)=ITMP
   10 CONTINUE
      RETURN
      END
\end{verbatim}
The function {\tt RANDOM()} is assumed to generate one
random real number uniformly distributed in the interval,
$I=\{ r; 0\leq r <1 \}$.

Above program is, however, dangerous.
We expect that the {\tt J} generated by {\tt J=INT(RANDOM()*DFLOAT(I+1))}
are distributed uniformly on $\{ 0$, $1$, $2$, $\cdots$, {\tt I}$\}$ but
it is not always correct, especially when {\tt I} is large.
This is because the random real numbers generated by a random number
generation routine take only discrete values.
For example, if the random real number takes one of the values in
$\{ 0$, $0.1$, $0.2$, $0.3$, $0.4$, $0.5$, $0.6$, $0.7$, $0.8$, $0.9\}$
uniformly and {\tt I=7}, the probabilities that {\tt J=0}, {\tt 1},
{\tt 2}, {\tt 3}, {\tt 4}, {\tt 5}, {\tt 6} and {\tt 7} are
$1/5$, $1/10$, $1/10$, $1/10$, $1/5$, $1/10$, $1/10$ and $1/10$, respectively.
Of course, if the random number generation routine generates
random real numbers which take all real*8 floating precision between
$0$ and $1$, such non-uniform property might be negligible.

The above mentioned problem is removed in the following routine
{\tt PERM2}:
\begin{verbatim}
      SUBROUTINE PERM2(IB,N,ILIST)
      IMPLICIT REAL*8 (A-H,O-Z)
      DIMENSION ILIST(0:N-1)
      IF((2**(IB-1).GE.N).OR.(2**IB.LT.N))THEN
        WRITE(*,*)'ERROR! IN PERM2. SPECIFIED IB=',IB, ' AND N=',N
        WRITE(*,*)'       ARE NOT IN PROPER RANGE.'
        STOP
      END IF
      DO 10 I=1,IB
        IN=2**(IB-I)
        IS=1-I
        IFIRST=IN*2-1
        IF(I.EQ.1)IFIRST=N-1
        DO 20 J=IFIRST,IN,-1
   30     K=ISHFT(IRNDTW(),IS)
          IF(K.GT.J)GOTO 30
c
          ITMP=ILIST(K)
          ILIST(K)=ILIST(J)
          ILIST(J)=ITMP
   20   CONTINUE
   10 CONTINUE
      RETURN
      END
\end{verbatim}
The function {\tt IRNDTW()} is assumed to generate random integers
between {\tt 0} to {\tt 2**IB-1} (See Appendix~A).
The values of {\tt IB} and {\tt N} must satisfy the relation
\begin{equation}
2^{\tt IB-1}<N\leq 2^{\tt IB}.
\end{equation}

Two different implementations of shuffling routines are given here.
The necessary number of random numbers for shuffling $N$ elements,
which is denoted by $n_R(N)$, are estimated here.
In the first subroutine {\tt PERM1},
$n_R(N)$ is $N$.
In the second subroutine {\tt PERM2}, $n_R(N)$ depends on the
values of generated random numbers. About one half of the generated random
numbers are not used.
In average, $n_R(N)$ is $2N$ but there is some fluctuation.
The following will be an overestimate of it:
\begin{equation}
         ( N         + 2 \sqrt{N}) +
         ({N\over 2} + 2 \sqrt{N\over 2}) +
         ({N\over 4} + 2 \sqrt{N\over 4}) + \cdots
       = 2 N + {2\sqrt{2}\over \sqrt{2} -1} \sqrt{N}.
\end{equation}
Therefore $n_R(N)=3N$ is an overestimated value and it is a safe
estimate.

The shuffling algorithm implemented in the {\tt PERM2} routine is
used in the following analysis.
And we can say that the random number cost for the shuffling of $N$
elements is $3N$ in the worst case and this $n_R(N)=3N$ is used in
the next section for the efficiency analysis of our recycle method.

\section{Efficiency}

The Monte Carlo estimation of an $M$-dimensional integration:
\begin{equation}
{\hat I}[F]=
\int_{x_1^{\rm min}}^{x_1^{\rm max}}
\int_{x_2^{\rm min}}^{x_2^{\rm max}}
\int_{x_3^{\rm min}}^{x_3^{\rm max}}
\cdots
\int_{x_M^{\rm min}}^{x_M^{\rm max}}
F(x_1, x_2, x_3, \cdots, x_M)
dx_1dx_2dx_3\cdots dx_M
\end{equation}
is considered to study the efficiency of our recycle method.
The dynamic Monte Carlo simulation will correspond to the
very large $M$ case.
After rescalings,
$y_i = (x_i - x_i^{\rm min})/(x_i^{\rm max}-x_i^{\rm min})$
$(i=1$, $2$, $3$, $\cdots$, $M)$, this integration reduces to the
estimation of
\begin{equation}
I[f] =
\int_0^1\int_0^1\int_0^1\cdots\int_0^1
f(y_1, y_2, y_3, \cdots, y_M)
dy_1dy_2dy_3\cdots dy_M.
\end{equation}
Instead of this integral, the discrete approximant $I^N_M[f]$ is
estimated by a Monte Carlo simulation. This $I^N_M[f]$ is expected to
converge to $I[f]$ when $N$ goes to infinity.

Before the Monte Carlo sampling is started,
$L_{\rm table}$ permutation tables are prepared and it requires
$n_R(N) L_{\rm table}$ random numbers.
The permutation operation in the $i$-th table is expressed by $s_i$
$(i=1$, $2$, $3$, $\cdots$, $L_{\rm table})$.
Then a set of $M$ random numbers $r_1$, $r_2$, $r_3$, $\cdots$, $r_M$
which is denoted by $\vec r\in \Omega^M$ is generated and the value of
$f(\vec r )$ is accumulated.
The average of $L_{\rm sample}$ samples is
denoted by $E_0(\{\vec r_i$; $i=1$, $2$, $3$, $\cdots$, $L_{\rm sample}\} )$.
At the same time, the values of
$f(s_i(r_1)$, $s_i(r_2)$, $s_i(r_3)$, $\cdots$, $s_i(r_M))$
$(i=1$, $2$, $3$, $\cdots$, $L_{\rm table})$ which is denoted by
$f(s_i(\vec r))$ are also accumulated.
The averages of $L_{\rm sample}$ samples are denoted by
$E_i(\{ \vec r_i\} )$.

Each of these estimations $E_i(\{ \vec r_j\} )$
$(i=0$, $1$, $2$, $\cdots$, $L_{\rm table})$
converges to $I^N_M[f]$ as is shown in lemma 3 and the error defined by
\begin{equation}
\delta_i (\{ \vec r_j\} ) = E_i(\{ \vec r_j\} )-I^N_M[f]
\end{equation}
converges as
\begin{equation}
<\delta_i (\{ \vec r_j\} )>_{
   \vec r_1, \vec r_2, \vec r_3, \cdots, \vec r_{L\rm sample} \in \Omega^M
 }
= {1\over L_{\rm sample}} I^N_M[\Delta f^2].
\end{equation}
Therefore the magnitude of the error is
\begin{equation}
|\delta_i(\{ \vec r_j\})| \sim \sqrt{I_M^N[\Delta f]\over L_{\rm sample}}.
\label{MCERROR}
\end{equation}
Now the behavior of the averages of $E_i(\{ \vec r_j\} )$ ($i=1$, $2$, $3$,
$\cdots$, $L_{\rm table}$), that is,
\begin{equation}
E_{\rm all}(\{ s_i\} , \{ \vec r_j\} ) = { 1\over L_{\rm table} }
   \sum_{i=1}^{L_{\rm table}} E_i (\{ \vec r_j\} )
\end{equation}
is studied.
It is obvious that this $E_{\rm all}$ converges to $I_M^N[f]$
when $L_{\rm sample}\rightarrow\infty$ even if $L_{\rm table}$ is finite.
But is it true when $L_{\rm table}$ goes to $\infty$ and $L_{\rm sample}$
is finite?
The question is how this quantity converges towards $I_M^N[f]$.
And the answer is: The error of the $E_{\rm all}$ defined by
\begin{equation}
\delta_{\rm all} (\{ s_i\} , \{ \vec r_j\} )
  = E_{\rm all}(\{ s_i\} , \{ \vec r_j\}) - I_M^N[f]
\end{equation}
behaves as
\begin{eqnarray}
& &\big< <
\delta_{\rm all}(\{ s_i\} , \{ \vec r_j\} )^2
>_{
   \vec r_1, \vec r_2, \vec r_3, \cdots, \vec r_{L_{\rm sample}}
   \in \Omega^M
}
\big>_{
s_1, s_2, s_3, \cdots, s_{L_{\rm table}}\in S_N
}
\nonumber\\
&=&
{1\over L_{\rm table} L_{\rm sample}}( I_M^N[\Delta f^2] +{C_{f,f}\over N} )
+{1\over L_{\rm sample}}{C_{f,f}\over N}
+O({1\over N^2}).
\label{RECYCLEFL}
\end{eqnarray}
It is concluded from this eq.~(\ref{RECYCLEFL})
that the error does not converge to zero when $L_{\rm table}
\rightarrow 0$ and $L_{\rm sample}$ is finite.
{}From eq.~(\ref{RECYCLEFL}), we can derive the condition under which
the Monte Carlo simulation with the recycle algorithm is advantageous to
the standard algorithm.

The number of necessary
random numbers to achieve a given error $\delta$ is denoted by
$\Lambda_S(\delta)$ in the standard algorithm case and $\Lambda_R(\delta )$
in the recycle algorithm case.
The $\Lambda_S(\delta )$ is obtained by solving the equation
the right-hand side of eq.~(\ref{MCERROR}) be $\delta$:
\begin{equation}
\sqrt{I_M^N[\Delta f^2]\over \Lambda_S(\delta )/M} =\delta.
\end{equation}
Therefore
\begin{equation}
\Lambda_S(\delta )={I_M^N[\Delta f^2 ] M\over \delta^2 }.
\end{equation}
To get the $\Lambda_R(\delta )$, we have to optimize the number of
tables, $L_{\rm table}$.
When we use $L_{\rm table}$ tables, the
accuracy is obtained from eq.~(\ref{RECYCLEFL}) and
\begin{equation}
\Lambda_R(\delta )=ML_{\rm sample} +n_R(N)L_{\rm table}
\end{equation}
When one random permutation table is generated, $n_R(N)$ random numbers are
necessary as studied in the previous section.
After we minimize the $\Lambda_R(\delta )$ under the condition that the
error be $\delta$, we reach the expression
\begin{equation}
\Lambda_R(\delta )={2\over \delta }\sqrt{
   n_R(N) M(I_M^N[\Delta f^2] +{C_{f,f}\over N})
}
+
{MC_{f,f}\over \delta^2 N}
\end{equation}
when
\begin{equation}
L_{\rm table} ={1\over \delta }\sqrt{{M\over n_R(N)}(I_M^N[\Delta
f^2]+{C_{f,f}\over
N})}.
\end{equation}
Therefore the recycle algorithm is efficient when
\begin{equation}
{I_M^N[\Delta f^2 ] M\over \delta^2 }
>
{2\over \delta }\sqrt{M n_R(N)(I_M^N[\Delta f^2]+{C_{f,f}\over N})}
+ {MC_{f,f}\over \delta^2 N}
\end{equation}
in terms of the number of random numbers.
Generally speaking, $I_M^N[\Delta f^2] /C_{f,f}$ is usually
much larger than one as is observed in the next section as an example
and $\delta\cdot N$ should be larger than one.
After neglecting irrelevant terms, above condition reduces to
\begin{equation}
\sqrt{MI_M^N[\Delta f^2]\over 4\delta^2 n_R(N)}>1,
\end{equation}
which is usually satisfied in the high accuracy simulation case.
Furthermore the optimal number of tables is
\begin{equation}
L_{\rm table} = \sqrt{MI_M^N[\Delta f^2] \over \delta^2 n_R(N)},
\end{equation}
which is a large number also in the high accuracy simulation case.

These efficiency studies here show that our recycle algorithm
in some cases is more efficient than the standard algorithm.
It is especially efficient for high accuracy Monte Carlo simulations.

Only the number of random numbers was considered to study the efficiency
in this section and there will be many other factors to determine
the computational time in the real computer simulation.
So we have to study the real efficiency and it is made in the next
section and the result proves that the recycle algorithm is
really efficient in high accuracy simulations.

\section{Simple Application}

In this section, an example is given to show that our recycle
algorithm really reduces the computational time without introducing
any bias to the estimate and the
lemma~4 and the recycle algorithm in section 2 works successfully.
The Monte Carlo estimation of the volume of an $M$-dimensional
hypersphere is considered in the following.

In this case, the integrand is
\begin{equation}
f(x_1, x_2, x_3, \cdots , x_M ) = \Big\lbrace
\begin{array}{cc}
2^M&\mbox{ if } \sum_{i=1}^M x_i^2 <1 \\
0&\mbox{ otherwise }\\
\end{array},
\end{equation}
where $0\leq x_i <1$($i=1$, $2$, $3$, $\cdots$, $M$).
The answer is
\begin{equation}
I[f] = V_M = {\pi^{M/2} \over \Gamma (M/2+1)},
\end{equation}
where $V_M$ denotes the volume of the $M$-dimensional unit hypersphere.
The constants relevant to the efficiency analysis,
$I_M^N[\Delta f^2]$ and $C_{f,f}$, are estimated to be
\begin{equation}
I_M^N[\Delta f^2] = (2^M -V_M )V_M
\end{equation}
and
\begin{equation}
C_{f,f} = {M(M-1)\over 2} (V_M - \sqrt{2}V_{M-1})^2.
\end{equation}

Two Monte Carlo programs are prepared for $M=5$ case.
In this case, $I_5^N [\Delta f^2] =(2^5 -V_5)V_5 \approx 140.7$
and $C_{f,f} =10(V_5 -\sqrt{2} V_4)^2 \approx 29.41$.
One does not use the recycle algorithm and the other one uses it.
They are optimized to their algorithms and the FORTRAN codes are
given in the appendix~A.
$64$ independent samples are used to estimate the statistical errors and
each sample is an average over $10^3\sim 10^7$ independent trials.
In the recycle algorithm case, $63$ random permutations are prepared
and used.

The results and performance for the $13$ bit-wide random integer case,
that is, the $N=2^{13}=8192$ case are given in Table~\ref{HS5D13}.
The exact value of the integral is $8\pi^2/15=5.26378901\cdots$.
The $I_{\rm est.}$ converges correctly except for the $N_{\rm trial}=10^7$
case.
The $N_{\rm trial}=10^7$ simulation is more accurate than the
accuracy limit determined by the random number bit-width.
In this calculation, the accuracy limit is of the order of
\begin{equation}
\delta_{\rm limit} = {2^M\over 2^B} = 2^{5-B},
\end{equation}
where $B$ is the bit-width of the random numbers.
The calculation reaches this accuracy after $N_{\rm limit}$
trials determined by
\begin{equation}
\sqrt{I_5^N[\Delta f^2]\over N_{\rm limit}} = \delta_{\rm limit},
\end{equation}
therefore
\begin{equation}
N_{\rm limit}
= {I_5^N[\Delta f^2]\over \delta_{\rm limit}^2 }
= {\pi^2 \over 15}(1-{\pi^2 \over 60})\cdot 2^{2 B -2}.
\end{equation}
When $B=13$, $\delta_{\rm limit} = 0.0039$ and
$N_{\rm limit} = 9.2\times 10^6$.
The above calculation with $N_{\rm trial} = 10^7$ means totally
$6.4\times 10^7$ trials which are much longer than $N_{\rm limit}$.
In the $N_{\rm trial}=10^6$ case, the estimated accuracy is of
the same order as the accuracy limit and therefore the results of
those cases are not reliable, although they seem to
be acceptable.
\begin{table}[p]
\begin{center}
\begin{tabular}{cccrcc}\hline\hline
$N_{\rm trial}$&algorithm&$I_{\rm est.}$&$X$&CPU time&$T_{\rm ratio}$\\\hline
$10^3$ & NR & $5.296  \pm 0.043$   & $ 0.75$& $0.8$ &      \\\cline{2-4}
       &  R & $5.217  \pm 0.040$   & $-1.17$& $2.8$ & $0.3$\\\hline
$10^4$ & NR & $5.254  \pm 0.015$   & $-0.65$&$6  $ &      \\\cline{2-4}
       &  R & $5.262  \pm 0.015$   & $-0.12$&$4  $ & $1.5$\\\hline
$10^5$ & NR & $5.2648 \pm 0.0057$  & $ 0.18$&$57 $ &      \\\cline{2-4}
       &  R & $5.2674 \pm 0.0044$  & $ 0.82$&$20 $ & $2.9$\\\hline
$10^6$ & NR & $5.2652 \pm 0.0014$  & $ 1.01$&$570$ &      \\\cline{2-4}
       &  R & $5.2654 \pm 0.0017$  & $ 0.95$&$179$ & $3.2$\\\hline
$10^7$ & NR & $5.26765\pm 0.00044$ & $ 8.77$&$5694$&      \\\cline{2-4}
       &  R & $5.26615\pm 0.00049$ & $ 4.82$&$1764$& $3.2$\\\hline\hline
\end{tabular}
\end{center}
\caption{The results and performance for $13$ bit-wide random number
case are shown.
The column named algorithm shows which program was used.
NR means the program with conventional random number usage and
R means the recycle use algorithm for random numbers.
The $N_{\rm trial}$, $I_{\rm est.}$ and $T_{\rm ratio}$
denote the number of trials for one sample, the estimated volume of the
five-dimensional unit hypersphere and the CPU time ratio of NR and R cases,
respectively.
$64$ samples are used to estimate the value of $I_{\rm est.}$ and its
error. The SUN Sparc Station 2 was used to measure the CPU time.
The $X$ denotes $(I_{\rm MC} - I_{\rm exact})/\delta I_{\rm MC}$,
where $I_{\rm MC}$, $I_{\rm exact}$ and $\delta I_{\rm MC}$
denote the estimated volume, exact volume and estimated error, respectively.
}
\label{HS5D13}
\end{table}

In Table~\ref{HS5D16}, the result and performance with $16$ bit-wide
random integers are shown.
In this case, the accuracy limit is $0.00049$ and the maximum number
of trials is $5.9\times 10^8$.
The calculations are tried up to $N_{\rm trial}=10^6$.
\begin{table}[p]
\begin{center}
\begin{tabular}{cccrcc}\hline\hline
$N_{\rm trial}$&algorithm&$I_{\rm est.}$&$X$&CPU time&$T_{\rm ratio}$\\\hline
$10^3$ & NR & $5.342  \pm 0.045$   & $ 1.74$& $0.8$ &      \\\cline{2-4}
       &  R & $5.238  \pm 0.051$   & $-0.51$& $ 21$ & 0.04 \\\hline
$10^4$ & NR & $5.249  \pm 0.015$   & $-0.99$& $  6$ &      \\\cline{2-4}
       &  R & $5.249  \pm 0.014$   & $-1.06$& $ 24$ & 0.25 \\\hline
$10^5$ & NR & $5.2665 \pm 0.0047$  & $ 0.58$& $ 58$ &      \\\cline{2-4}
       &  R & $5.2602 \pm 0.0053$  & $-0.68$& $ 52$ & 1.1  \\\hline
$10^6$ & NR & $5.2619 \pm 0.0016$  & $ 1.18$& $572$ &      \\\cline{2-4}
       &  R & $5.2634 \pm 0.0014$  & $-0.28$& $340$ & 1.7  \\\hline\hline
\end{tabular}
\end{center}
\caption{
The results and performance for $16$ bit-wide random number case are shown.
The notations are the same as those in Table.~1.
}
\label{HS5D16}
\end{table}

The results in Tables~\ref{HS5D13} and \ref{HS5D16} show two things clearly:
(1)our recycle algorithm works correctly and (2) it accelerates the
high-precision and long Monte Carlo simulations.

\section{Ising Monte Carlo Simulation}

An example which uses the recycle algorithm
naively was shown in the previous section
and an advanced application is given in this section.
It is the Ising Monte Carlo simulation with an independent-system
coding technique.

The most efficient algorithm of this problem prepares and uses
tables named {\it Boltzmann-factor tables}\cite{CM86,NIYK88,KIK92}
which convert the value of a random integer into a few bits
variable which is useful to update the spins coded in one computer
word. For example, in the case of the ferromagnetic Ising model on a cubic
lattice with nearest neighbor interaction without external field,
two tables {\tt IX1} and {\tt IX2} are used.
The $l$th bits of these tables express the value of two-bit
variables for the system coded in the $l$th bit.
When the spins in one word are updated, one random integer {\tt IR}
is converted into two values {\tt IX1(IR)} and {\tt IX2(IR)}.
Each pair of bits in the same bit-positions of these {\tt IX1(IR)}
and {\tt IX2(IR)} holds sufficient information about the random number
to update each Ising spin coded in that bit-position.
They are used to update the spins and {\tt IR} is not necessary
anymore.
The recycle algorithm can be implemented by shuffling randomly
the bit pairs
in {\tt IX1} and {\tt IX2} before the spin updating operations.
The subroutine {\tt BFSHUF} for this operation is given in Appendix~B
and the shuffling operation is described in this FORTRAN subroutine.
The algorithm is essentially the same as {\tt PERM2} routine given
in the third section.
The performance of this {\tt BFSHUF} routine is measured with
the NEC SX3/11 in K\"oln university and the MONTE4\cite{MFOUR}
in Japan Atomic Energy Research Institute,
and it is given in Table~\ref{PERFBFSHUF}.
The MONTE4 has four processors and the performance is measured
with one processor of them under single job environment, which is
appropriate to compare the performance with the single processor
machine, NEC SX3/11.
The performance of the MONTE4 is largely influenced by the other
jobs on other processors and the performance shown in
Table~\ref{PERFBFSHUF} suffers from large fluctuations
under multi-job environment.
Although the time consuming operation of this routine is not
vectorizable, the Boltzmann factor tables are shuffled within
reasonable CPU time.
\begin{table}[p]
\begin{center}
\begin{tabular}{ccc}\hline\hline
{\tt IRBIT} & SX3/11 & MONTE4 \\\hline
$10$ & $0.046$ sec. & $0.039$ sec. \\
$11$ & $0.091$  & $0.078$ \\
$12$ & $0.19$ & $0.17$   \\
$13$ & $0.43$ & $0.37$   \\
$14$ & $1.0$  & $0.85$    \\
$15$ & $2.2$  & $2.0$    \\
$16$ & $4.7$  & $4.2$    \\
$17$ & $9.8$  & $8.4$     \\
$18$ & $ 20$  & $17$   \\
$19$ & $ 41$  & $37$   \\
$20$ & $ 82$  & $74$  \\
$21$ & $165$  & $149$  \\
$22$ & $332$  & $284$  \\
$23$ &        & $568$  \\\hline\hline
\end{tabular}
\end{center}
\caption{The shuffling performance of the {\tt BFSHUF} routine
given in the Appendix B is listed.
The {\tt IRBIT} denotes the bit-width of the random integer and
the size of the Boltzmann factor tables to be shuffled is $2^{\tt IRBIT}$.
The $8$ byte integer is specified for integer variables and arrays
with {\it -w} option of the NEC FORTRAN77/SX compiler.
The CPU time for the shuffling is measured and shown. For small value
of {\tt IRBIT}, the {\tt BFSHUF} routine is called repeatedly so that
the time may be the order of ten seconds and the averaged CPU time is
shown.
The {\tt IRBIT}$=23$ performance of SX3/11 is not available
because of the memory-size limitation. }
\label{PERFBFSHUF}
\end{table}
This Ising Monte Carlo algorithm with table shuffling is useful because
the systems at the same temperature can be simulated with the
independent-system coding technique.

Fig.~\ref{CORPLOT} shows two sequences of the magnetization of $64$
systems at the
critical temperature and no correlation is observed in this
figure.
The correlation between the magnetization sequences is studied
quantitatively here to check the validity of our algorithm.
The magnetizations from the $i$th permuted random sequences are
denoted by
\begin{equation}
M_i(t), \quad t=1, 2, 3, \cdots, m,
\end{equation}
where $t$ denotes the sample number in the temporal order
and the magnetization is calculated in every constant Monte Carlo steps.
We assume that
there are $n$ sequences, that is, $i=1$, $2$, $3$, $\cdots$, $n$.
The correlation coefficient between samples $i$ and $j$ is defined by
\begin{equation}
c_{ij} ={<M_i(t) M_j(t)>_t \over \sqrt{<M_i(t)^2>_t <M_j(t)^2>_t}},
\end{equation}
where $<\cdot >_t$ denotes the temporal average, that is,
\begin{equation}
<f(t)>_t = {1\over m} \sum_{t=1}^m f(t).
\end{equation}
The average of $c_{ij}$ over all different samples $i$ and $j$,
\begin{equation}
c={\sum_{i\neq j} c_{ij} \over \sum_{i\neq j} 1 }
\end{equation}
is estimated for several simulations with different parameters.
The results are given in Table~\ref{MCORRELATION} and it is
clearly observed that the correlation between the samples using permuted
random numbers is negligibly small as is expected from our lemma~4.
\begin{figure}[t]

\vspace{10cm}

\caption{The values of magnetizations $M_1(t)$ and $M_2(t)$ of two
samples with different permutations are plotted.
After $10^4$ MCS, the magnetizations were calculated $100$ times at every
$1000$ MCS at the critical temperature.
Five independent simulations were made and therefore there are $500$ points.
No clear correlation is observed.
The quantitative analysis of the correlation is given in Table~4.
}
\label{CORPLOT}
\end{figure}
\begin{table}[p]
\begin{center}
\begin{tabular}{c|cc|cc|cc}\hline\hline
       &\ \ \ \ \ {\tt IRBIT}&$=16$&\ \ \ \ \ {\tt IRBIT}&$=18$
&\ \ \ \ \ {\tt IRBIT}&$=20$ \\
  $m$  &$c$&$X$          &$c$&$X$          &$c$&$X$\\\hline
$10^2$ &$-0.0011$ & $-0.92$&$ 0.0002 $ & $0.17$&$-0.0009$ &$-0.75$\\
$10^3$ &$ 0.00003$& $0.08$&$-0.00007$ & $-0.18$&$ 0.00028$&$0.72$ \\
$10^4$ &$ 0.00016$& $1.33$&$ 0.00012$ & $1.00$&$ 0.00011$&$0.92$ \\
$10^5$ &$ 0.000014$&$0.39$&$0.000052$ & $1.44$&$0.000053$&$1.36$ \\\hline\hline
\end{tabular}
\end{center}
\caption{The correlation $c$ between the magnetization sequences from the
simulation with permuted random numbers is shown.
$X$ denotes $c/\delta c$, where $\delta c$ denotes the estimated error of $c$.
The system on a $11\times 11\times 12$ lattice was simulated at
the critical temperature.
The initial $10^4$ MCS is skipped.
The $m$ denotes the number of samples and the magnetization is
measured every $50$ MCS.
Five independent simulations are used to estimate the errors.
No correlation is observed in these results.
}
\label{MCORRELATION}
\end{table}

\section{Conclusion and Remarks}

It is shown that the random permutation or random
shuffling of the numbers in the interval $[0$, $1]$ transforms one
uniformly distributed random sequences into another statistically
almost independent random sequences.
Based on this property, a new method named {\it recycle algorithm}
for random number generation and its usage is proposed.
Its efficiency is studied. It is also shown that this algorithm
reduces the computational time of long and high-precision
Monte Carlo simulation.
This algorithm is especially useful in the Ising Monte Carlo
simulation with independent-system coding technique.
It makes it clear that the random shuffling of the Boltzmann
factor tables\cite{CM86} makes
the statistically independent simulation possible for the systems
with the same parameters using that coding technique.

This recycle algorithm will be also useful in Monte Carlo
simulations with MIMD architecture parallel computer.
One processor generates random integers and broadcasts them to
all the other processors. Each processor holds one random permutation
and transforms the broadcasted random integers with its table.
When the cost for broadcasting is smaller than the generation,
this method will be useful.

The automatic optimization of the bit-width of the random numbers
and the number of permutation tables is possible by using the
results of efficiency analysis in the fourth section.
It will be useful for moderate-scale Monte Carlo simulation.

In the very large scale dynamics Monte Carlo simulation, this
algorithm is often useful. The number of tables and bit-width
can be determined by the necessary accuracy of the simulation and
memory capacity of the computer.

\section*{Acknowledgments}

The authors thank D.~Stauffer for stimulating discussion
and SFB341 for partial support.
The calculations were performed on the NEC SX3/11 and the SUN workstations of
K\"oln university and the MONTE4 of Japan Atomic Energy Research
Institute.

\section*{Appendix A--Program for Hypersphere}

The FORTRAN programs for the Monte Carlo integration of the volume
of five-dimensional hypersphere used in the sixth section are shown.
Two kinds of programs are listed in the following.
One uses the conventional algorithm and the other uses the
recycle algorithm.

\subsection*{Conventional Random Number Use}

The simulation parameters are specified in two {\tt PARAMETER} statements
at the beginning.
{\tt NSIZE} and {\tt ISAMPL} specify the number of trials in one sample
and the number of samples, respectively.
{\tt IRBIT} and {\tt IRSEED} specify the bit-width of the random number
and the initial seed.
{\tt IRSEED} should be an odd integer.
The random number generation routines are given at the end of
this Appendix~A.

{\small
\begin{verbatim}
      IMPLICIT REAL*8 (A-H,O-Z)
      PARAMETER(NSIZE=10000000,ISAMPL=64)
      PARAMETER(IRBIT=16,IRSEED=14643557)
      DIMENSION ACC(ISAMPL)
C INITIALIZATION OF RANDOM NUMBER GENERATION ROUTINE
      CALL INITTW(IRSEED,IRBIT)
C MONTE CARLO SAMPLING
      DO 10 I=1,ISAMPL
        ACC(I)=SPHAMR(NSIZE,IRBIT)
 10   CONTINUE
C ERROR ANALYSIS
      AAVE=0.0D0
      ASQ=0.0D0
      DO 20 I=1,ISAMPL
        AAVE=AAVE+ACC(I)
        ASQ=ASQ+ACC(I)*ACC(I)
 20   CONTINUE
      AAVE=AAVE/DFLOAT(ISAMPL)
      ASQ=ASQ/DFLOAT(ISAMPL)
      AERR=DSQRT((ASQ-AAVE*AAVE)/DFLOAT(ISAMPL-1))
C OUTPUT
      WRITE(6,100)NSIZE,ISAMPL,(NSIZE*ISAMPL)
 100  FORMAT(' TRIAL=',I8,' SAMPLE=',I6,' TOTAL TRIAL=',I10)
      WRITE(6,110)IRBIT,2.0d0**(-IRBIT)
 110  FORMAT(' BIT WIDTH=',I2,'  RNG RESOLUTION=',F10.8)
      WRITE(6,120)AAVE,AERR
 120  FORMAT(' ESTIMATED VOLUME=',F10.8,'+-',F10.8)
      STOP
      END
      FUNCTION SPHAMR(N,IRBIT)
      IMPLICIT REAL*8 (A-H,O-Z)
      ANORMS=4.0D0**IRBIT
      ICNT=0
      DO 10 I=1,N
        A1=DFLOAT(IRNDTW())
        A2=DFLOAT(IRNDTW())
        A3=DFLOAT(IRNDTW())
        A4=DFLOAT(IRNDTW())
        A5=DFLOAT(IRNDTW())
        A=A1*A1+A2*A2+A3*A3+A4*A4+A5*A5
        IF(A.LT.ANORMS)ICNT=ICNT+1
 10   CONTINUE
      SPHAMR=DFLOAT(ICNT)*32.0D0/DFLOAT(N)
      RETURN
      END
\end{verbatim}
}

\subsection*{Recycle Algorithm}

In the following program, the number of trials of one sample is
{\tt NSIZE} $\times$ {\tt IREP}. {\tt ISAMPL} specifies the
number of random permutation tables and the original
random sequence is also used.
Therefore the number of sample is {\tt ISAMPL}$+1$.
The random number generation routine at the end of this
Appendix and the subroutine {\tt PERM2} in the third section are also
necessary.

{\small
\begin{verbatim}
      IMPLICIT REAL*8 (A-H,O-Z)
      PARAMETER(IRBIT=16,NSIZE=1000000,IREP=10,ISAMPL=63)
      PARAMETER(IRSEED=14643557)
      PARAMETER(ILIST=2**IRBIT)
      DIMENSION IX(0:ILIST-1,ISAMPL),IR(NSIZE,5)
      DIMENSION ACC(0:ISAMPL)
      CALL INITTW(IRSEED,IRBIT)
      DO 10 J=1,ISAMPL
        DO 20 I=0,ILIST-1
          IX(I,J)=I
 20     CONTINUE
        CALL PERM2(IRBIT,ILIST,IX(0,J))
 10   CONTINUE
      DO 30 I=0,ISAMPL
        ACC(I)=0.0D0
 30   CONTINUE
      DO 40 I=1,IREP
        CALL ITWDIM(NSIZE*5,IR)
        ACC(0)=ACC(0)+SPHAMC(NSIZE,IR(1,1),IR(1,2),
     1         IR(1,3),IR(1,4),IR(1,5),IRBIT)
        DO 50 J=1,ISAMPL
          ACC(J)=ACC(J)+SPHAMD(NSIZE,IR(1,1),IR(1,2),
     1           IR(1,3),IR(1,4),IR(1,5),IRBIT,ILIST,IX(0,J))
 50     CONTINUE
 40   CONTINUE
      AAVE=0.0D0
      ASQ=0.0D0
      DO 70 I=0,ISAMPL
        ACC(I)=ACC(I)/DFLOAT(IREP)
        AAVE=AAVE+ACC(I)
        ASQ=ASQ+ACC(I)*ACC(I)
 70   CONTINUE
      AAVE=AAVE/DFLOAT(ISAMPL+1)
      ASQ=ASQ/DFLOAT(ISAMPL+1)
      AERR=DSQRT((ASQ-AAVE*AAVE)/DFLOAT(ISAMPL))
C OUTPUT
      WRITE(6,100)NSIZE*IREP,ISAMPL+1,NSIZE*IREP*(ISAMPL+1)
 100  FORMAT(' TRIAL=',I8,' SAMPLE=',I6,' TOTAL TRIAL=',I10)
      WRITE(6,110)IRBIT,2.0d0**(-IRBIT)
 110  FORMAT(' BIT WIDTH=',I2,'  RNG RESOLUTION=',F10.8)
      WRITE(6,120)AAVE,AERR
 120  FORMAT(' ESTIMATED VOLUME=',F10.8,'+-',F10.8)
      STOP
      END
      FUNCTION SPHAMC(N,IR1,IR2,IR3,IR4,IR5,IRBIT)
      IMPLICIT REAL*8 (A-H,O-Z)
      DIMENSION IR1(N),IR2(N),IR3(N),IR4(N),IR5(N)
      ANORMS=2.0D0**IRBIT
      ANORMS=ANORMS*ANORMS
      ICNT=0
      DO 10 I=1,N
        A1=DFLOAT(IR1(I))
        A2=DFLOAT(IR2(I))
        A3=DFLOAT(IR3(I))
        A4=DFLOAT(IR4(I))
        A5=DFLOAT(IR5(I))
        A=A1*A1+A2*A2+A3*A3+A4*A4+A5*A5
        IF(A.LT.ANORMS)ICNT=ICNT+1
 10   CONTINUE
      SPHAMC=DFLOAT(ICNT)*32.0D0/DFLOAT(N)
      RETURN
      END
      FUNCTION SPHAMD(N,IR1,IR2,IR3,IR4,IR5,IRBIT,M,IX)
      IMPLICIT REAL*8 (A-H,O-Z)
      DIMENSION IR1(N),IR2(N),IR3(N),IR4(N),IR5(N)
      DIMENSION IX(0:M-1)
      ANORMS=2.0D0**IRBIT
      ANORMS=ANORMS*ANORMS
      ICNT=0
      DO 10 I=1,N
        A1=DFLOAT(IX(IR1(I)))
        A2=DFLOAT(IX(IR2(I)))
        A3=DFLOAT(IX(IR3(I)))
        A4=DFLOAT(IX(IR4(I)))
        A5=DFLOAT(IX(IR5(I)))
        A=A1*A1+A2*A2+A3*A3+A4*A4+A5*A5
        IF(A.LT.ANORMS)ICNT=ICNT+1
 10   CONTINUE
      SPHAMD=DFLOAT(ICNT)*32.0D0/DFLOAT(N)
      RETURN
      END
\end{verbatim}
}

\subsection*{Random Number Generation Routine}

The random number generation routine used in the above
two programs is given.
The Tausworthe sequence (or Kirkpatrick and Stoll method)
is used to generate the pseudo-random integers.

{\small
\begin{verbatim}
      SUBROUTINE INILSD(N)
      COMMON /LCSEED/NSEED
      NSEED=N
      RETURN
      END
      FUNCTION LCARNG()
      INTEGER*4 LCARNG
      DATA MASK31/Z'7FFFFFFF'/
      COMMON /LCSEED/NSEED
      NSEED=NSEED*48828125
      NSEED=IAND(NSEED,MASK31)
      LCARNG=NSEED
      RETURN
      END
      SUBROUTINE INITTW(IRSEED,IRBIT)
      PARAMETER(ISELBT=-24)
      COMMON /TWSTOR/IDIM(250),IPL(250),IQL(250),IP
      DATA IRCNST/Z'7FFFFFFF'/
      CALL INILSD(IRSEED)
      DO 10 I=1,250
         IDIM(I)=0
         DO 20 J=1,32
           ITMP=LCARNG()
           ITMP=ISHFT(ITMP,ISELBT)
           ITMP=IAND(ITMP,1)
           IDIM(I)=ISHFT(IDIM(I),1)
           IDIM(I)=IOR(IDIM(I),ITMP)
   20    CONTINUE
   10 CONTINUE
C BIT MASKING
      IRLST=0
      DO 40 I=1,IRBIT
        IRLST=ISHFT(IRLST,1)
        IRLST=IOR(IRLST,1)
   40 CONTINUE
      DO 30 I=1,250
        IDIM(I)=IAND(IDIM(I),IRLST)
   30 CONTINUE
      IP=1
      DO 50 I=1,250
        IPL(I)=I+1
        IT=I-103
        IF(IT.LT.1)IT=IT+250
        IQL(I)=IT
 50   CONTINUE
      IPL(250)=1
      RETURN
      END
      FUNCTION IRNDTW()
      COMMON /TWSTOR/IR(250),IPL(250),IQL(250),IP
      IRNDTW=IEOR(IR(IP),IR(IQL(IP)))
      IR(IP)=IRNDTW
      IP=IPL(IP)
      RETURN
      END
      SUBROUTINE ITWDIM(N,IDIM)
      COMMON /TWSTOR/IR(250),IPL(250),IQL(250),IP
      DIMENSION IDIM(N)
      DO 10 I=1,N
        IDIM(I)=IRNDTW()
 10   CONTINUE
      RETURN
      END
\end{verbatim}
}


\section*{Appendix B--Shuffling of Boltzmann-factor Table}

The shuffling routine for Boltzmann factor tables(BFT) of Ising
Monte Carlo simulation with independent-system coding technique
is given.
This BFSHUF routine is for the simulation with two BFT and it is easily
modified for more BFT simulation.
This routine uses vectorized efficient random-number-generation
routine {\tt RNDO2I} of {\tt RNDTIK} library\cite{IK90}
which generates Kirkpatrick-Stoll type
random integers\cite{KS80}.
{\tt RNDO2I} is assumed to be initialized to generate {\tt IRBIT}-bit
wide random integers.

{\small
\begin{verbatim}
C   1992.10.26. BY NOBUYASU ITO
C               BOLTZMANN FACTOR TABLE SHUFFLING ROUTINE
C   IX1, IX2 : BOLTZMANN FACTOR TABLE TO BE SHUFFLED
C   IRBIT    : BIT-WIDTH OF THE RANDOM NUMBER SHOULD BE GIVEN
C          THE SIZE OF IX1 AND IX2 IS ASSUMED TO BE
C          IX1(0:2**IRBIT-1) IX2(0:2**IRBIT-1).
C   IRD(N)   : WORK AREA. ANY SIZE N IS GOOD BUT LARGE VALUE
C          (MORE THAN SEVERAL THOUSANDS) IS MORE EFFICIENT.
      SUBROUTINE BFSHUF(IX1,IX2,IRBIT,N,IRD)
      PARAMETER(IWIDTH=64,NMAX=30)
      DIMENSION IX1(0:2**IRBIT-1),IX2(0:2**IRBIT-1)
      DIMENSION IRD(N)
      IRPOS=1
      CALL RNDO2I(N,IRD)
      DO 10 I=0,IWIDTH-2
        IPOS=ISHFT(1,I)
        DO 20 J=1,IRBIT
          IN=2**(IRBIT-J)
          IS=1-J
          DO 30 K=IN*2-1,IN,-1
            DO 40 L=1,NMAX
              IR=ISHFT(IRD(IRPOS),IS)
              IRPOS=IRPOS+1
              IF(IRPOS.GT.N)THEN
                CALL RNDO2I(N,IRD)
                IRPOS=1
              END IF
              IF(IR.LE.K)GOTO 50
 40         CONTINUE
            WRITE(*,*)'ERROR IN BFSHUF! RANDOM NUMBER IS BIASED.'
            STOP
 50         ID=IEOR(IX1(K),IX1(IR))
            ID=IAND(ID,IPOS)
            IX1(K)=IEOR(IX1(K),ID)
            IX1(IR)=IEOR(IX1(IR),ID)
            ID=IEOR(IX2(K),IX2(IR))
            ID=IAND(ID,IPOS)
            IX2(K)=IEOR(IX2(K),ID)
            IX2(IR)=IEOR(IX2(IR),ID)
 30       CONTINUE
 20     CONTINUE
 10   CONTINUE
      RETURN
      END
\end{verbatim}
}


\begin{thebibliography}{99}
\bibitem[1]{NISMA}N.~Ito and M.~Suzuki, {\it J.~Phys.~Soc.~Jpn.} {\bf 60},
 1978(1991).
\bibitem[2]{NISMB}N.~Ito, {\it AIP Conf. Proc.} {\bf 248} {\it
Computer-Aided Statistical Physics} ed. C.-K.~Hu (AIP, 1992) p.~136.
\bibitem[3]{LF91}A.~M.~Ferrenberg and D.~P.~Landau, {\it Phys.~Rev.} {\bf B44},
 5081(1991).
\bibitem[4]{BGHP92}C.~F.~Baille, R.~Gupta, K.~A.~Hawick and G.~S.~Pawley,
{\it Phys.~Rev.} {\bf B45}, 10438 (1992).
\bibitem[5]{BDS86}G.~Bhanot, D.~Duke and R.~Salvador,
{\it Phys.~Rev.} {\bf B33}, 7841 (1986).
\bibitem[6]{CM86}C.~Michael, {\it Phys.~Rev.} {\bf B33}, 7861 (1986).
\bibitem[7]{NIYK88}N.~Ito and Y.~Kanada, {\it Supercomputer} {\bf 5} No.~3, 31
(1988).
\bibitem[8]{NIYK90}N.~Ito and Y.~Kanada, Proc. of {\it Supercomputing '90}
(New York, Nov. 1990) p.753.
\bibitem[9]{H90}H.-O.~Heuer, {\it Comp.~Phys.~Comm.} {\bf 59}, 387 (1990).
\bibitem[10]{R92}H.~Rieger, {\it Fast Vectorized Algorithm for the
Monte Carlo Simulation of the Random Field Ising Model},
to be published in {\it J.~Stat.~Phys.} (Jan. 1993).
\bibitem[11]{KIK92}N.~Kawashima, N.~Ito and Y.~Kanada,
{\it Algorithms for Monte Carlo Simulations of the Ising Models on
a Simple Cubic Lattice},
to be published in {\it Int.~J.~Mod.~Phys.} {\bf C}.

\bibitem[12]{KO86}M.~Kikuchi and Y.~Okabe, {\it Prog.~Theor.~Phys.} {\bf 75},
192 (1986).
\bibitem[13]{ISPTP87}N.~Ito and M.~Suzuki, {\it Prog.~Theor.~Phys.} {\bf 77},
1391 (1987).
\bibitem[14]{KO87PTP}M.~Kikuchi and Y.~Okabe, {\it Prog.~Theor.~Phys.} {\bf
78},
540 (1987).
\bibitem[15]{KO87PR}M.~Kikuchi and Y.~Okabe, {\it Phys.~Rev.} {\bf B35},
5382 (1987).
\bibitem[16]{IS88}N.~Ito and M.~Suzuki, {\it J.~Physique~Col. {\bf C8} Supp.},
1565 (1988).
\bibitem[17]{HEUPR90}H.-O.~Heuer, {\it Phys.~Rev.}~{\bf B42}, 6476 (1990).
\bibitem[18]{HEUEP90}H.-O.~Heuer, {\it Europhys.~Lett.}~{\bf 12}, 551 (1990).
\bibitem[19]{ISPROC91}N.~Ito and M.~Suzuki in {\it Computer Simulation Studies
in Condensed Matter Physics III}, {\it Springer Proc. Phys.} {\bf 53} ed. by
D.~P.~Landau, K.~K.~Mon and H.-B. Sch\"utter (Springer-Verlag, 1991) p.16.
\bibitem[20]{ISPRB91}N.~Ito and M.~Suzuki, {\it Phys.~Rev.}~{\bf B43},
3483 (1991).
\bibitem[21]{HEUJP92}H.-O.~Heuer, {\it J.~Phys.}~{\bf A25}, L567 (1992).
\bibitem[22]{NINISHI}N.~Kawashima, N.~Ito and M.~Suzuki, {\it
J.~Phys.~Soc.~Jpn.}
{\bf 61}, 1777 (1992).
\bibitem[23]{NIPHY92}N.~Ito, {\it Non-equilibrium Critical Relaxation
of the Three-dimensional Ising Model},
to be published in {\it Physica} {\bf A}.

\bibitem[24]{KO92}M.~Kikuchi and Y.~Okabe, {\it J.~Phys.~Soc.~Jpn.} {\bf 61}
No.~10 (1992).
\bibitem[25]{TAUS65}R.~C.~Tausworthe, {\it Math.~Comput.} {\bf 19}, 201(1965).
\bibitem[26]{LP73}T.~S.~Lewis and W.~H.~Payne, {\it J.~ACM} {\bf 20}, 456
(1973).
\bibitem[27]{KS80}S.~Kirkpatrick and E.~P.~Stoll, {\it J.~Comp.~Phys.} {\bf
40},
517 (1980).
\bibitem[28]{IK90}N.~Ito and Y.~Kanada, {\it Supercomputer} {\bf 7} No.~1, 29
(1990).
\bibitem[29]{MFOUR}K.~Asai, K.~Higuchi, M.~Akimoto, H.~Matsumoto and Y.~Seo,
{\it JAERI Monte Carlo Machine}, to be published in the Proc.~of
{\it Mathematical Methods and Supercomputing in Nuclear Applications}
(Karlsruhe, 1993, April).

\end{thebibliography}
\end{document}